\documentclass[10 pt, journal]{IEEEtran}

\usepackage{graphicx}
 \usepackage{verbatim}
 \usepackage{amsmath}
 \usepackage{eepic}
\usepackage[ruled,vlined]{algorithm2e}
\usepackage{amssymb}

\newcommand{\mb}[1]{\mathbb{#1}}
\newcommand{\bs}[1]{\boldsymbol{#1}}
\newcommand{\mc}[1]{\mathcal{#1}}

\newtheorem{definition}{Definition}

\newcommand{\zJT}[1]{z_{#1}(s_{#1-1}, s_{#1})}
\newcommand{\hJT}[1]{h_{#1}(s_{#1-1},s_{#1})}

\newcommand{\uSum}[1]{\Big(\bigoplus_{\bs s} u(\bs s)\Big)^{(#1)}}

\begin{document}


\title{Entropy Semiring Forward-backward Algorithm \\ for HMM Entropy Computation}

\author{Velimir M. Ili\'c
\thanks{V. Ili\'c is with the Department of Informatics, Faculty of Sciences and Mathematics, University of Ni\v s,
Serbia.}
}

\maketitle

\begin{abstract}
The paper presents Entropy Semiring Forward-backward algorithm
(\emph{ESRFB}) and its application for memory efficient
computation of the subsequence constrained entropy and state
sequence entropy of a Hidden Markov Model (\emph{HMM}) when an
observation sequence is given. \emph{ESRFB} is based on
forward-backward recursion over the entropy semiring, having the
lower memory requirement than the algorithm developed by Mann and
MacCallum, with the same time complexity. Furthermore, when it is
used with forward pass only, it is applicable for the computation
of \emph{HMM} entropy for a given observation sequence, with the
same time and memory complexity as the previously developed
algorithm by Hernando et al.

\end{abstract}

\section{Introduction}

Hidden Markov Models (\emph{HMMs}) are standard probabilistic
models for state sequences in sequential data labeling
\cite{Rabiner_89}. Subsequence constrained entropy of \emph{HMM}
explaining an observation sequence and state sequence entropy, are
useful quantities which provide a measure of \emph{HMM}
uncertainty. One criterion for the estimation of the \emph{HMM}
quality is the entropy of state sequence explaining an observation
sequence, which provides a measure of its uncertainty
\cite{Mann_McCallum_07}, \cite{Hernando_et_al_05}.

The algorithms for \emph{HMMs} mostly consider efficient
marginalization which is usually performed using the
forward-backward algorithm (\cite{Khreich_et_al_10}), which runs
in $\mc O (N^2 T)$ time, where $N$ denotes the number of states
and $T$ is the length of sequence. Recently, Mann and MacCllum
have developed an algorithm for computation of \emph{HMM}
subsequence constrained entropy for similar probabilistic model
conditional random fields (\emph{CRF}), which is based on the
marginal probabilities computation \cite{Mann_McCallum_07} with
the same asymptotical complexity as $FB$. This algorithm can be
adapted to work with \emph{HMMs}, but when the sequence length is
large it becomes memory demanding, since it needs $\mc O(NT)$
memory. On the other hand, Hernando et al.
\cite{Hernando_et_al_05} developed the memory efficient algorithm
for state sequence entropy computation which requires $\mc O(N)$
memory. The algorithm has the same time complexity as \emph{FB},
but it is not applicable for the computation of subsequence
constrained entropy.

In this paper we develop a new algorithm which can be used for
both types of computations. The algorithm is based on
forward-backward recursion over the entropy semiring
\cite{Ilic_et_al_11} and is called Entropy Semiring
Forward-backward algorithm (\emph{ESRFB}). \emph{ESRFB} has lower
memory requirement than Mann-MacCallum's algorithm subsequence
constrained entropy computation. Furthermore, when it is used with
the forward pass only it can compute the entropy in the same time
and space as Hernando et al.'s algorithm. Moreover, it is shown
how the Hernando et al.'s algorithm can be derived from
\emph{ESRFB}.

The paper is organized as follows. In section II we define the
\emph{HMM} and present the forward-backward algorithm (\emph{FB})
for efficient marginalization of \emph{HMM}. Section III reviews
the algorithms by Hernando et. al. and Mann and McCallum, for
efficient computation of \emph{HMM} entropy and subsequence
constrained entropy for a given observation sequence. Section IV
gives the general \emph{FB} algorithm which operates over the
commutative semiring. Finally, section V considers the \emph{FB}
over the entropy semiring and its application to \emph{HMM}
entropy computation.

\section{Hidden Markov Models and forward-backward algorithm}
\label{hmm}


In this paper, we adopt the following notation:

\begin{itemize}

\item The sequence $l, l+1, \dots, r$ is shortly denoted with $l:r$, and the sequence $0:l, r:T$ is denoted
with $- l:r $

\item Big letters are used for random variables ($S_t, O_t$) and the small ones for their realizations ($s_t, o_t$).

\item The sequence of symbols is $(S_l, \dots S_r)$ is denoted with $S_{l:r}$, the sequence $(S_0, \dots, S_l, S_r, \dots, S_T)$ with $S_{-l:r}$ and similarly for $s_{l:r}$, $O_{l:r}$ and
$o_{l:r}$ for $0 \leq l \leq r \leq T$.

\item The sequences $S_{0:T}$, $O_{0:T}$, $s_{0:T}$ and $o_{0:T}$ are denoted with $\bs S$, $\bs O$, $\bs s$ and $\bs o$, respectively.

\item The variables are omitted in probability notation. Thus, $p(s_t, o_{1:t})$ stands for $P(S_t=s_t, O_{1:t} =o_{1:t})$, $p(\bs o)$ for $P(\bs O =\bs o)$ and so on.

\end{itemize}

Hidden Markov model (\textit{HMM}) consists of the following
elements:

\begin{itemize}

\item A Markov chain $(S_0, \dots , S_T)$, represented by an $N \times N$ stochastic
matrix $A$, which describes the transition probabilities
$a_{ij} = P(S_t = j | S_{t-1} = i)$ between the $N$ states of the model, together
with a probability distribution $b_i$, where $\pi_i = P(S_0 = i)$.

\item A set of probability distributions, one for each hidden state,
$b_i(o_t) = P(O_t = o_t | S_t = i)$, which model the emission of
such observations. If there are $M$ possible distinct
observations, we accommodate the probability distributions to be
in the rows of an $N \times M$ matrix $B$.

\end{itemize}

With these settings, the joint probability that state sequence
$\bs S$  takes value $\bs s$ and the observation sequence  $\bs O$
takes value $\bs o$ is given with:
\begin{equation}
\label{HMM p expanded}
p(\bs s, \bs o)=\pi_{s_0} b_{s_0}(o_0) \prod_{t=1}^T a_{s_{t-1} s_t} b_{s_t}(o_t).
\end{equation}

Using the probability conditions $\sum_{s_t} a_{s_{t-1}, s_t} = 1$
and $\sum_{o_t} b_{s_t} (o_t) =1$, we can derive two important
equations which characterizes \emph{HMM} and will be used in the
rest of the paper:
\begin{align}
\label{hmm: p_subseq}%
&p(s_{0:i}, o_{0:i}) =%
\pi_{s_0} b_{s_0}(o_0)%
\prod_{t=1}^i a_{s_{t-1} s_t} b_{s_t}(o_t),\\
\label{hmm: p_subseq_right}
&p(s_{i+1:T}, o_{i+1:T} | s_t) =%
\prod_{t=i+1}^T a_{s_{t-1} s_t} b_{s_t}(o_t).
\end{align}


One of the main problems in \textit{HMMs} is efficient marginalization of computation of the \textit{HMM} conditional probability $p(\bs s | \bs o)$:
\begin{equation}
\label{hmm: p_lr}
p(s_{l:r} | \bs o )=\sum_{s_{-l:r}} p(\bs s | \bs o ) =\sum_{s_{-l:r}} \frac{p(\bs s, \bs o)}{p(\bs o)}.
\end{equation}

The computation of (\ref{hmm: p_lr}) 
by enumerating all the $\bs s \in \mc S^{T+1}$ requires about $T
N^{T+1}$  additions and multiplications, which would be infeasible
even for small values of $N$ and $T$ (for $N=10$ and $T=20$, the
total number of operations has an order $10^{22}$). A more
efficient way is the \textit{forward-backward}(\textit{FB})
algorithm which solves the problems by use of $\mc O(N^2 T)$
operations. In this paper we present a numerical stable variant of
\textit{FB}. For another variants see \cite{Rabiner_89},
\cite{Bishop_06}, \cite{Khreich_et_al_10}

The forward-backward algorithm recursively computes desired
quantities using the \emph{HMM} forward and backward probabilities:
\begin{equation}
\label{hmm: alpha_def}
\hat{\alpha_t}(s_t) 
= p(s_t| o_{1:t}), \quad
\hat{\beta}_t(s_t) = \frac{p(o_{t+1:T} | s_t)}{p(o_{t+1:T}|o_{0:t})},
\end{equation}
as follows.

\subsubsection{Forward initialization} For $1\leq j \leq N$:

\begin{equation}
c_0 = \sum_{j=1}^N \pi_{j} b_{j}(o_0),\quad
\hat \alpha_0(j) = \frac{\pi_{j} b_{j}(o_0)}{c_0},
\end{equation}

\subsubsection{Forward recursion} 
For $0\leq t \leq T$, $1\leq j \leq N$:
\begin{align}
c_t &=\sum_{j=1}^N \sum_{i=1}^N \alpha_{t-1}(i) a_{i j} b_{j}(o_{t}),\\
\hat{\alpha}_{t}(j)&=\frac{\sum_{i=1}^N \alpha_{t-1}(i) a_{i j} b_{j}(o_{t})}{c_t},
\end{align}
\subsubsection{Backward initialization} For $1\leq i \leq N$:
\begin{align}
&\hat\beta_T (i) = 1, 
\end{align}
\subsubsection{Backward recursion} For $T-1\geq t \geq 0$, $1\leq i,j \leq N$:
\begin{align}
\hat{\beta}_{t}(i)&=\frac{\sum_{j=1}^N a_{i j}b_{j}(o_{t+1}) \hat\beta_{t+1}(j)}{c_{t+1}}. 
\end{align}

The normalization factors $c_t$ ensure that the probabilities sums
to one and represents the conditional observational probabilities:
\begin{equation}
\label{hmm: c0_ct}
c_0=p(o_0),\quad c_t=p(o_t | o_{0:t-1}).
\end{equation}
Once the forward and backward probabilities are computed we can
compute the marginal as
\begin{equation}
\label{hmm: p_lr_solution}
p(s_{l:r}|\bs o)=
\alpha_l(s_l) \cdot \prod_{t=l+1}^r \frac{a_{s_{t-1}, s_t}{b_{s_t}(o_t)}}{c_t}  \cdot \beta_r(s_r)
\end{equation}
Two most commonly used marginals $p(s_{t-1:t} | \bs o)$ and $p(s_t | \bs o)$ 
can be computed as follows
\begin{align}
\label{hmm: xi_t}
p(s_{t-1:t} | \bs o)
&= \frac{\hat{\alpha}_{t-1}(s_{t-1})a_{s_{t-1} s_{t}}b_{s_{t}}(o_{t})\hat{\beta}_{t}(o_t)}{c_t},\\
\label{hmm: gamma_t}
p(s_t | \bs o) &=
\hat{\alpha}_{t}(s_{t}) \cdot \hat{\beta}_{t}(s_{t}).
\end{align}

The majority of computations are performed in the forward and
backward recursion phases, which results in the time complexity
$\mc O (N^2 T)$. The storing of all forward and backward vectors
along with the normalization factors requires $\mc O(NT)$ memory.

\section{Entropy computation of Hidden Markov models}

The conditional entropy of \textit{HMM} is given with
\begin{equation}
\label{hmm_en: H_S_o}
H(\bs S\ |\ \bs o) =
-\sum_{\bs s} p(\bs s \ | \ \bs o) \log p(\bs s \ |\ \bs o).
\end{equation}
while, the subsequence constrained entropy is
\begin{multline}
\label{hmm_en: H_S_lr_def}
H(S_{- l:r} | s_{l:r}, o_{0:T}) = \\
-\sum_{s_{- l:r}}
p(s_{- l:r} | s_{l:r}, o_{0:T}) \cdot
\log p(s_{- l:r} | s_{l:r}, o_{0:T}).
\end{multline}
If we introduce
\begin{equation}
\label{hmm_en: H_S_lr_tmp}
H(S_{-l:r} , s_{l:r}| \bs o) =
-\sum_{s_{- l:r}}
p(\bs s | \bs o) \cdot
\log p(\bs s |  \bs o),
\end{equation}
we can derive the following equality
\begin{equation}
\label{hmm_en: H_S_lr_final}
H(S_{- l:r} | s_{l:r},  \bs o) = \frac{H(S_{- l:r} , s_{l:r}| \bs o) + \log p(s_{l:r} | \bs o )}{p(s_{l:r} | \bs o )}
\end{equation}

A direct evaluation of (\ref{hmm_en: H_S_o}) is infeasible as
there are $N^T$ terms. In the following text we consider efficient
algorithms for the entropy computation.

First, in the next two subsections, we review two algorithms based
on the entropy decomposition rules \cite{Cover_Thomas_06}
\begin{align}
\label{hmm_en:H_dec}
H(X, Y ) &= H(X) + H(Y |X), \\
H(Y |X) &= \sum_{x} p(x)\cdot H(Y |X = x).
\end{align}
After that, in the next section we derive the new algorithms based on the \textit{ESRFB} algorithm.

\subsection{The algorithm by Mann and McCallum}

Mann and McCallum proposed the algorithm for the linear chain
\textit{conditional random fields} entropy gradient computation
\cite{Mann_McCallum_07}, which can also be used for \textit{HMMs}.
The algorithm uses the conditional probabilities
\begin{align}
\label{hmm_en: fbe: p_t_t+1}
\hat p_{t | t+1}(i|j)&= p(s_t | s_{t+1}, \bs o)=\frac{p(s_{t:t+1} | \bs o)}{p(s_{t+1}| \bs o)}, \\
\label{hmm_en: fbe: p_t_t-1}
\hat p_{t | t-1}(i|j)&= p(s_t | s_{t-1}, \bs o)=\frac{p(s_{t-1:t} | \bs o)}{p(s_{t-1}| \bs o)},
\end{align}
which, in turn, are computed using the \textit{FB} algorithm and
the forward and backward entropies, which, in turn, are computed
with the recursive procedure based on the entropy decomposition
formulas (\ref{hmm_en:H_dec}). The forward entropy
$H_t^{\alpha}(s_t)$ at time $t$ is defined as the entropy of state
sequence $S_{0:t-1}$ which ends in $s_t$, for a given observation
sequence $\bs o$:
\begin{equation}
H_t^{\alpha}(s_t) =
H(S_{0:t-1} | s_{t}, \bs o).
\end{equation}
while the backward entropy $H_t^{\beta}(s_t)$ at time $t$ is the entropy of state sequence $S_{t+1:T}$ which starts in $s_t$:
\begin{equation}
H_t^{\beta}(s_t) =
H(S_{t+1:T} | s_{t}, \bs o).
\end{equation}
Using the forward and backward entropies, subsequence constrained
entropy conditional \textit{HMM} entropy can be recursively
computed as in the following algorithm.

\subsubsection{Forward backward algorithm} Compute and store forward and backward probabilities using \textit{FB} algorithm.

\subsubsection{Forward entropy initialization} For $1 \leq j \leq N$:
\begin{equation}
H_0^\alpha(j)=0;
\end{equation}

\subsubsection{Forward entropy recursion} for $0\leq t \leq T-1$, $1 \leq i,j \leq N$:
\begin{align}
\label{H Fw recursive}
H_{t+1}^{\alpha}&(j)= %
\sum_{i=1}^N p_{t | t+1}(i|j)\Big( H_t^{\alpha}(i) -\log p_{t | t+1}(i|j)\Big),
\end{align}
where $\hat p_{t | t+1}(i|j)$ is computed using (\ref{hmm_en: fbe: p_t_t+1}), (\ref{hmm: xi_t}) and (\ref{hmm: gamma_t}).

\subsubsection{Backward entropy initialization} For $1 \leq j \leq N$:
\begin{equation}
H_T^\beta(j)=0;
\end{equation}

\subsubsection{Backward entropy recursion} for $0\leq t \leq T-1$, $1 \leq i,j \leq N$:
\begin{align}
\label{H Fw recursive}
H_{t-1}^{\beta}&(j)= %
\sum_{i=1}^T p_{t | t-1}(i|j)\Big( H_t^{\beta}(i) -\log p_{t | t-1}(i|j)\Big).
\end{align}
where $\hat p_{t | t+1}(i|j)$ is computed using (\ref{hmm_en: fbe: p_t_t-1}), (\ref{hmm: xi_t}) and (\ref{hmm: gamma_t}).

\subsubsection{Termination}
\begin{align}
H(&S_{- l:r}, s_{l:r}| \bs o) =\nonumber\\
&=p(s_{l:r} | \bs o)(H_l^{\alpha}(s_l) +H_r^{\beta}(s_r) +\log p(s_{l:r}| \bs o)),\\
H(&S_{- l:r} | s_{l:r},  \bs o) = \frac{H(S_{- l:r} , s_{l:r}| \bs o) + \log p(s_{l:r} | \bs o )}{p(s_{l:r} | \bs o )}
\end{align}

The time complexity of algorithm is $\mc O(N^2 T + N^{r-l})$,
where $\mc O(N^2 T)$ is for the forward-backward entropy
computation and $\mc O(N^{r-l})$ for the termination phase. The
memory complexity depends on the sequence length since all forward
and backward vectors should be available in forward and backward
entropy recursion phases; regarding $\mc O(N^{r-l})$ space
required for storing the results in the termination phase, the
total memory complexity is $O(N T + N^{r-l})$.

The algorithm can also be used for the computation of entropy
using the equality
\begin{equation}
H(\bs S | \bs o) = H(S_T | \bs o) + \sum_{s_{0:T-1}} p(s_T | \bs o) \cdot H_T^{(\alpha)}(s_T),
\end{equation}
which follows from the entropy decomposition formulas and
definition of forward entropy. In this case, the backward entropy
pass is not needed, but the time and memory complexity are not
reduced, since the forward and backward probabilities still need
to be computed. In the following subsection we review the
algorithm developed in \cite{Hernando_et_al_05} by Hernando et
al., which computes the entropy with the memory complexity
independent of the sequence length.

\subsection{The algorithm by Hernando et al.}

In \cite{Hernando_et_al_05}, Herando et al. develop the recursive
algorithm for the computation of Hidden Markov model entropy. It
uses \emph{HMM} forward probability
\begin{equation}
\hat \alpha_t(s_t)=p(s_t|o_{1:t}),
\end{equation}
conditional probability
\begin{equation}
\hat{p}_{t|t-1}(s_{t} | s_{t-1})=p(s_{t-1}|s_t,o_{1:t}),
\end{equation}
and intermediate entropy
\begin{equation}
H_t(s_t)=H(S_{0:t-1}|s_t,o_{1:t}).
\end{equation}
\textit{HMM} entropy is computed as follows.

\vskip 0.5cm

\subsubsection{Initialization}
For $1 \leq j \leq N$ set:
\begin{equation}
\label{hmm: en: H_0}%
H_0(j)=0,
\end{equation}
\begin{equation}
\label{hmm: en: alpha_0}%
\hat \alpha_0(j)=\frac{\pi_j b_j(o_0)}{\sum_{i=1}^N \pi_i b_i(o_1)}.
\end{equation}


\subsubsection{Induction}
For $1 \leq t \leq T$ and $1 \leq j \leq N$ set:
\begin{equation}
\label{hmm: en: alpha_t}%
\hat \alpha_t(j)=%
\frac{ \sum_{i=1}^N \hat \alpha_{t-1}(i) a_{ij} b_j(o_t) }
{\sum_{k=1}^N \sum_{i=1}^N \hat \alpha_{t-1}(i)a_{ik}b_k(o_t)},
\end{equation}
\begin{equation}
\label{hmm: en: p_t1_t}%
p_{t-1|t}(i|j)=\frac{\hat \alpha_{t-1}(i) a_{ij}}{\sum_{k=1}^N
\sum_{i=1}^N \hat \alpha_{t-1}(k) a_{kj}},
\end{equation}
\begin{multline}
\label{hmm: en: H_t}%
H_t(j)=\sum_{i=1}^N p_{t-1|t}(i|j)\Big(H_{t-1}(i)- \log p_{t-1|t}(i|j)\Big).
\end{multline}

\subsubsection{Termination}

\begin{equation}
\label{hmm: en: H_S_o}
H(\bs S\ |\ \bs o)=\sum_{j=1}^N
\hat \alpha_T(j)\Big(H_T(j)-\log \hat \alpha_T(j)\Big).
\end{equation}

The algorithm runs with the linear time complexity $O(N^2 T)$ and
in fixed memory space independent of sequence length, $O(N^2)$,
since the vectors $\hat \alpha_{t-1}$, $H_{t-1}$ and the matrix
$p_{t-1|t}$ should be computed only once in $t - 1$-th iteration
and, after having been used for the computation of $H_{t}$, they
can be deleted.

\section {The forward-backward over the commutative semiring}

The \textit{FB} algorithm for \textit{HMMs} works for more general
models in which the factors in (\ref{HMM p expanded}) are not
probabilities but the functions whose range is a commutative
semiring \cite{Aji_McEliece_00}. In this section we present the
forward-backward over the commutative semiring and derive the
\textit{FB} for \textit{HMMs} as a special case.

\subsection{The forward-backward algorithm over a commutative semiring}

We begin with the definition of the commutative semiring.

\begin{definition}
A \emph{commutative semiring} is a set $\mathbb K$
with operations $\oplus$ and $\otimes$ such that both $\oplus$
and $\otimes$ are commutative and associative and have identity
elements in $\mathbb K$ ($\overline 0$ and $\overline 1$ respectively), and $\otimes$ is
distributive over $\oplus$.
\end{definition}

Let $\bs s=\big\{s_0, \dots, s_T\big\}$ be a set of variables taking values from the set $\mc S = \{1, \dots, N \}$.  
We define the \textit{local kernel} functions, $u_0: \mc S
\rightarrow \mathbb K$, $u_t: \mc S^2 \rightarrow \mathbb K$ for
$t=1, \dots , T$, and the \textit{global kernel} function $u: \mc
S^{T+1} \rightarrow \mathbb K$, assuming that the following
factorization holds
\begin{equation}
\label{fsr: u-chain}%
u(\bs s)= u_0(s_0) \otimes \bigotimes_{t=1}^{T} u_t(s_{t-1}, s_{t})
\end{equation}
for all $\bs s = (s_0, \dots, s_T)  \in \mc S^{T+1}$.

The \textit{FB} algorithm solves two problems

\begin{enumerate}

\item The \textit{marginalization problem:} Compute the sum
\begin{equation}
\label{fsr: v_ab}
v_{a:b}(s_{a:b})=\bigoplus_{s_{0:T - a:b}} u(\bs s),
\end{equation}

\item The \textit{normalization problem:} Compute the sum
\begin{equation}
\label{fsr: Z}
Z = \bigoplus_{\bs s} u(\bs s).
\end{equation}

\end{enumerate}

Similarly as in \textit{HMM}, the \textit{FB} recursively computes
the \textit{forward variable}
\begin{equation}
\label{fsr: alpha_def}
\alpha_i(s_i)=\bigoplus_{s_{0:i-1}}%
u_0(s_0) \bigotimes_{t=1}^{i} u_t(s_{t-1}, s_{t}),
\end{equation}
which is initialized to
\begin{equation}
\label{fsr: alpha_init} \alpha_0(s_0)=u_0(s_0),
\end{equation}
and recursively computed using
\begin{equation}
\label{fsr: alpha_rec}%
\alpha_i(s_i)=\bigoplus_{s_{i-1}}
u_{i-1}(s_{i-1}, s_i) \otimes \alpha_{i-1}(s_{i-1}),
\end{equation}%
and the \textit{backward variable}
\begin{equation}
\label{fsr: beta_def}
\beta_i(s_i)=\bigoplus_{s_{i+1:T}}\ %
\bigotimes_{t=i+1}^{T} u_t(s_{t-1}, s_{t}),
\end{equation}
which is recursively computed using
\begin{equation}
\beta_t(s_t)=\bigoplus_{s_{t+1}} u_{t+1}(s_t, s_{t+1}) \otimes \beta_{t+1}(s_{t+1}),
\end{equation}
and initialized to
\begin{equation}
\beta_T(s_T)=1.
\end{equation}
Once, the forward $\alpha_l$ and backward $\beta_r$ variables are computed, we can solve the marginalization problem by use of the formula
\begin{equation}
\label{fsr: v_ab}
v_{l:r}(s_{l:r})=
\alpha_l(s_l) \otimes
\bigotimes_{i=l+1}^r u_i(s_{i-1}, s_i) \otimes \beta_r(s_r).
\end{equation}

The normalization problem can be solved with the forward pass only according to
\begin{equation}
\label{FB Norm Final}%
\bigoplus_{\bs s} u(\bs s)=
\bigoplus_{s_{T}} \alpha_{T}(s_{T}).
\end{equation}

In the following subsection we derive the \textit{FB} algorithm for \textit{HMMs} as a special case of the \textit{FB} over the commutative semiring.

\subsection{HMM forward-backward as a special case of the forward-backward over the commutative semiring}
\label{fsr: hmm}

\newcommand{\abHMM}[3]{a_{#1 #2} b_#2(o_#3)}
\newcommand{\aHMM}[2]{a_{#1 #2}}
\newcommand{\bHMM}[2]{b_#1(o_#2)}

The conditional \textit{HMM} probability $p(\bs s | \bs o)$ can be
seen as a special case of the global kernel factorization
(\ref{fsr: u-chain}) if $\oplus$ and $\otimes$ stand for the
addition and multiplication of the real numbers. To clarify this,
recall that join \textit{HMM} probability (\ref{HMM p expanded})
has the form
\begin{equation}
\label{fsr: hmm: p_join}
p(\bs s, \bs o)=\pi_{s_0} b_{s_0}(o_0) \prod_{t=1}^T a_{s_{t-1} s_t} b_{s_t}(o_t),
\end{equation}
and that according to the chain rule, conditional observational probability can be represented as
\begin{equation}
p(o_{0:T})=p(o_0) \cdot \prod_{t=1}^T p(o_t | o_{0:t-1})=
c_0 \cdot \prod_{t=1}^T c_t,
\end{equation}
where $c_0=p(o_0)$ and $c_t=p(o_t | o_{0:t-1})$ as in (\ref{hmm: c0_ct}). Then,
\begin{equation}
\label{fsr: hmm: p_s_o_fact}
p(\bs s | \bs o)=\frac{p(\bs s , \bs o)}{p(\bs o)}=z_0(s_0) \prod_{t=1}^T \zJT{t},
\end{equation}
where
\begin{equation}
z_0(s_0)=\frac{\pi_{s_0} b_{s_0}(o_0)}{c_0},\quad
z_t(s_{t-1},s_t)= \frac{a_{s_{t-1} s_t} b_{s_t}(o_t)}{c_t}.
\end{equation}
According to the equation (\ref{hmm: p_subseq}), the subsequence
conditional probabilities can be represented as
$p(s_{0:i} | o_{0:i}) = z_0(s_0) \prod_{t=1}^i \zJT{t}$,
and the forward variable (\ref{fsr: alpha_def}) has the form
\begin{equation}
\label{fsr: alpha_def}
\alpha_i(s_i)=\sum_{s_{0:i-1}}%
z_0(s_0) \prod_{i=t}^{i} z_t(s_{t-1}, s_{t})=p(s_{i} | o_{0:i}),
\end{equation}
in agreement with (\ref{hmm: alpha_def}). The recursive equations
(\ref{fsr: alpha_rec}), (\ref{fsr: alpha_init}) for the forward
variable have the form
\begin{equation}
\alpha_0(s_0)= z_0(s_0) = \frac{\pi_{s_0} b_{s_0}(o_0)}{c_0},
\end{equation}
\begin{multline}
\alpha_t(s_t) = \sum_{s_{t-1}} \zJT{t}\cdot \alpha_{t-1}(s_{t-1}) = \\
=\frac{\sum_{s_{t-1}} a_{s_{t-1} s_t} b_{s_{t}}(o_{t}) \alpha_{t-1}(s_{t-1})}{c_t},
\end{multline}
and the normalization factors can be computed using the probability condition
\begin{equation}
\sum_{s_t} \alpha_t(s_t) = \sum_{s_t} p(s_{t} | o_{0:t}) = 1,
\end{equation}
which gives
%
\begin{equation}
c_0= \sum_{j=1}^N \pi_{j} b_{j}(o_0) \quad
c_t=\sum_{j=1}^N \sum_{i=1}^N \alpha_{t-1}^{(z)}(i)  \abHMM{i}{j}{t}.
\end{equation}
and the definition of forward variable and its recursive equations agrees with the equations from section \ref{hmm}. 
Similarly, according to the equation (\ref{hmm: p_subseq_right}),
$p(s_{i+1:T}, o_{i+1:T} | s_t)= \prod_{t=i+1}^{T} z_t(s_{t-1},s_{t})$,
the backward variable is
\begin{equation}
\label{fsr: beta_def}
\beta_i(s_i)=\sum_{s_{i+1:T}}%
\prod_{t=i+1}^{T} z_t(s_{t-1}, s_{t})=
\frac{p(o_{t+1:T} | s_t)}{p(o_{t+1:T}|o_{0:t})},
\end{equation}
with the recursive equation:
\begin{equation}
\beta_T(s_T)=1,
\end{equation}
\begin{equation}
\beta_t(s_t) =\frac{\sum_{s_{t+1}} a_{s_{t} s_{t+1}} b_{s_t}(o_t) \alpha_{t-1}(s_{t-1})}{c_{t+1}}.
\end{equation}
Finally, in the same manner, the equation (\ref{fsr: v_ab}) reduces to (\ref{hmm: p_lr_solution})
\begin{equation}
\label{hmm: p_lr_solution}
p(s_{l:r}|\bs o)=
\alpha_l(s_l) \cdot \prod_{t=l+1}^r \frac{a_{s_{t-1},
s_t}{b_{s_t}(o_t)}}{c_t}  \cdot \beta_r(s_r),
\end{equation}
which retrives the \textit{HMM} forward-backward algorithm.

\section {The forward-backward algorithm over the entropy semiring}
\label{esrfa}

In this section we consider the forward-backward algorithm over
the entropy semiring (\emph{ESRFB}) and its application to
\textit{HMM} entropy computation. The entropy semiring
(\emph{ESR}), which is introduced in \cite{Cortes_et_al_08} and
\cite{Eisner_02}, is defined as follows.

\begin{definition}
\label{esrfa: esr_operations_def} The entropy semiring is a the commutative semiring for which $\mb K =\mb R^2$ and the semiring operations are defined with:
\begin{eqnarray}
\label{ESR plus}
(z_1, h_1) \oplus (z_2, h_2)  &=& (z_1+z_2, h_1+ h_2), \\
\label{ESR times}
(z_1, h_1) \otimes (z_2, h_2) &=& %
(z_1 z_2 ,\ z_1 h_2 + z_2 h_1),
\end{eqnarray}
for all $(z_1, h_1)$, $(z_2,  h_2)$ from $\mb R^2$. The identities
for $\oplus$ and $\otimes$ are $(0,0)$ and $(1,0)$, respectively.
\end{definition}

The first component of an ordered pair is called a $z$-part, while
the second one is an $h$-part. The following lemma can be proven
by the induction (see \cite{Ilic_et_al_11}).

\newtheorem{lemma}{Lemma}

\begin{lemma}
\label{esrfa: lema: 1}
Let $(z_i, z_i h_i)\in \mc R^2$ for all $0 \leq i \leq T$. Then,
the following equality holds:

\begin{equation}
\label{esrfa: lemma}
\bigotimes_{i=0}^{T}(z_i,z_i h_i)=%
\biggl(\ \prod_{i=0}^{T} z_i \ ,%
\ \prod_{i=0}^{T} z_i \ \cdot  \sum_{j=0}^{T} h_j \ \biggl).
\end{equation}
\end{lemma}

Let the local kernels in (\ref{fsr: u-chain}) have the form:
\begin{align}
\label{JT_ESR_factor_zero}
u_0(s_0) &= \big( z_0(s_0), z_0(s_0)h_0(s_0)\big);\\
\label{JT_ESR_factor_t}
u_i(s_{i-1}, s_i) &= \big( z_i(s_{i-1}, s_i), z_i(s_{i-1}, s_i) h_i(s_{i-1}, s_i)\big),
\end{align}
where
\begin{equation}
\label{emp_hmm: z_0}
z_0(s_0)=\frac{\pi_{s_0} b_{s_0}(o_0)}{c_0},\quad
z_t(s_{t-1},s_t)= \frac{a_{s_{t-1} s_t}
b_{s_t}(o_t)}{c_t}.
\end{equation}
with $c_0=p(o_0)$, $c_t=p(o_t | o_{0:t-1})$ and
\begin{align}
h_0(s_0)= \log z_0(s_0),\quad
\hJT{t}=\log \zJT{t}.
\end{align}
From Lemma 1, it follows that the $z$ and $h$ parts of the global kernel
\begin{equation}
\label{JT esr u prod}%
u(\bs s)= u_0(s_0) \otimes \bigotimes_{i=1}^{T} u_i(s_{i-1}, s_i),
\end{equation}
are given with:
\begin{align}
\label{esrfa: u_global_z}
u(\bs s)^{(z)} = z_0(s_0) &\prod_{i=1}^T z_i(s_{i-1}, s_i)\\
\label{esrfa: u_global_h}
u(\bs s)^{(h)} =
z_0(s_0) &\prod_{i=1}^T z_i(s_{i-1}, s_i)\cdot \\
&\big(h_0(s_0) + \sum_{j=1}^T h_j(s_{j-1}, s_j)\big).
\end{align}
Note that
\begin{equation}
h_0(s_0)+\sum_{j=1}^T \hJT{j} = \log \big(z_0 \cdot \prod_{j=1}^T \zJT{j}\big),
\end{equation}
and, according to the factorization (\ref{fsr: hmm: p_s_o_fact})
for \textit{HMM} conditional probability
$p(\bs s | \bs o)= z_0 \cdot \prod_{i=1}^T \zJT{i}$,
we can represent the global kernel as follows
\begin{equation}
\label{esrfa: u_p_plogp} u(\bs s) = \big(\ p(\bs s | \bs o)\ , \
p(\bs s | \bs o)\log p(\bs s | \bs o) \ \big).
\end{equation}
Hence, by summing of the global kernel we can obtain the entropies
$H(\bs S | \bs o)$ or $H(S_{-l:r} , s_{l:r}| \bs o)$ as the $h$
part of the sum, which depends on the set of the variables which
are summed out. Two types of the summation correspond to the
normalization and marginalization of the global kernel which can
be solved with the forward-backward algorithm over the entropy
semiring.

The $z$ and $h$ parts of the forward and backward variables in the
entropy semiring can also be derived using Lemma \ref{esrfa: lema:
1}. For the forward vector,
\begin{equation}
\alpha_t(s_t)=%
\bigoplus_{s_{0:t-1}} u_0(s_0) \otimes \bigotimes_{i=1}^{t} u_i(s_{i-1}, s_i)
\end{equation}
we have
\begin{align}
\label{esrfa: alpha_z_def}
\alpha_t^{(z)}(s_t)=%
\sum_{s_{0:t-1}}
z_0(s_0) \cdot &\prod_{i=1}^{t} z_i(s_{i-1}, s_i) \\
\label{esrfa: alpha_h_def}
\alpha_t^{(h)}(s_t)=%
\sum_{s_{0:t-1}}\ z_0(s_0) \cdot &\prod_{i=1}^t z_i(s_{i-1}, s_i)\cdot \\
&\big(h_0(s_0) + \sum_{j=1}^t h_j(s_{j-1}, s_j)\big),
\end{align}
and by use of the equality
$p(s_{0:t} | o_{0:t}) = z_0(s_0) \prod_{i=1}^t \zJT{i}$,
%
we obtain
\begin{align}
\alpha_t^{(z)}(s_t) &=%
\sum_{s_{o:t}} p(s_{0:t} | o_{0:t}) = p(s_t | o_{0:t}),\\
\alpha_t^{(h)}(s_t) &=%
\sum_{s_{o:t}} p(s_{0:t} | o_{0:t}) \log p(s_{0:t} | o_{0:t}).
\end{align}
The $z$-part of the \emph{ESR} forward vector is the \emph{HMM}
forward probability as defined in the sections \ref{hmm} and
\ref{fsr: hmm}, while the information about subsequence entropies
is propagated through the $h$-part, so that at each step we have
\begin{equation}
H(S_{0:t} | o_{0:t})= \sum_{s_t} \alpha_t^{(h)}(s_t).
\end{equation}

The forward vector is initialized to $u_0(s_0)$ and regarding
(\ref{JT_ESR_factor_zero}) we have:
\begin{align}
\label{emp_hmm: alpha_z_init}
\alpha_0^{(z)}(s_0)=z_0(s_0) & =\frac{\pi_{s_o} b_{s_o}(o_0)}{c_0}, \\
\label{emp_hmm: alpha_h_init}%
\alpha_0^{(h)}(y_0)=z_0(s_0)& h_0(s_0)=\nonumber \\%
&=\frac{\pi_{s_o} b_{s_o}(o_0)}{c_0}%
\log \frac{\pi_{s_o} b_{s_o}(o_0)}{c_0}.
\end{align}

The $z$ and $h$ forward recursive equation
\begin{equation}
\alpha_i(s_i)=\bigoplus_{s_{i-1}}
u_{i-1}(s_{i-1}, s_i) \otimes \alpha_{i-1}(s_{i-1}),
\end{equation}%
can be determined using the definition of the entropy semiring as
\begin{align}
\alpha_i^{(z)}(s_i)=&\sum_{s_{i-1}} \zJT{i} \alpha_{i-1}^{(z)}(s_{i-1}) \\
\alpha_i^{(h)}(s_i)=&\sum_{s_{i-1}} \zJT{i} \ \cdot \nonumber\\
&\big(\alpha_{i-1}^{(h)}(s_{i-1}) + \hJT{i}
\alpha_{i-1}^{(z)}(s_{i-1})\big),
\end{align}
or equivalently
\begin{align}
\label{emp_hmm: alpha_z_rec}
\alpha_t^{(z)}&(j)=\sum_{i=1}^N \frac{a_{ij} b_j(o_t)}{c_t} \cdot \alpha_{t-1}^{(z)}(i) \\
\label{emp_hmm: alpha_h_rec}
\alpha_i^{(h)}&(j)=\sum_{i=1}^N \frac{a_{ij} b_j(o_t)}{c_t} \ \cdot \nonumber\\
&\big(\alpha_{t-1}^{(h)}(i) + \alpha_{t-1}^{(z)}(i) \log
\frac{a_{ij} b_j(o_t)}{c_t} \big).
\end{align}
Similarly as in sections \ref{hmm} and \ref{fsr: hmm}, factors
$c_t$ can be found by normalization of $z$-parts:
\begin{equation}
c_0=\sum_{j=1}^N \pi_{j} b_{j}(o_0)\quad
\label{emp_hmm: c_t}
c_t =\sum_{j=1}^N \sum_{i=1}^N%
\alpha_{t-1}^{(z)}(i) \abHMM{i}{j}{t}.
\end{equation}

The backward vector
\begin{equation}
\label{emp_hmm: beta_def}
\beta_i(s_i)=\bigoplus_{s_{i+1:T}}\ %
\bigotimes_{t=i+1}^{T} u_t(s_{t-1}, s_{t}),
\end{equation}
has corresponding $z$ and $h$ parts
\begin{align}
\beta_t^{(z)}(s_t)&=%
\sum_{s_{t+1:T}}
\prod_{i=t+1}^{T} z_i(s_{i-1}, s_i) \\
\beta_t^{(h)}(s_t)&=%
\sum_{s_{t+1:T}}\prod_{i=t+1}^T z_i(s_{i-1}, s_i)\cdot
\sum_{j=t+1}^T h_j(s_{j-1}, s_j).
\end{align}
The equality (\ref{hmm: p_subseq_right}) implies
\begin{equation}
\prod_{i=t+1}^T z_i(s_{i-1}, s_i)=%
\frac{p(s_{t+1:T}, o_{t+1:T} | s_t)}{p(o_{t+1:T}|o_{0:t})}%
\end{equation}
and we have
\begin{align}
&\beta_t^{(z)}(s_t)=%
\frac{p(o_{t+1:T} | s_t)}{p(o_{t+1:T}|o_{0:t})} \\
&\beta_t^{(h)}(s_t)= \nonumber \\ %
&\sum_{s_{t+1:T}} %
\frac{p(s_{t+1:T}, o_{t+1:T} | s_t)}{p(o_{t+1:T}|o_{0:t})} %
\log \frac{p(s_{t+1:T}, o_{t+1:T} |s_t)}{p(o_{t+1:T}|o_{0:t})},
\end{align}
which gives the $z$-part of the \emph{ESR} backward vector, the
same as \emph{HMM} backward probability from the sections
\ref{hmm} and \ref{fsr: hmm}.

The backward vector is initialized according to $\beta_T(s_T)=1$:
\begin{equation}
\label{emp_hmm: beta_z_init}
\beta_T(s_T)^{(z)}=1,\quad
\beta_T(s_T)^{(h)}=0.
\end{equation}

while the recursive equation
\begin{equation}
\beta_t(s_t)=\bigoplus_{s_{t+1}} u_{t+1}(s_t, s_{t+1}) \otimes \beta_{t+1}(s_{t+1}),
\end{equation}
reduces to
\begin{align}
\beta_i^{(z)}(s_i)=&\sum_{s_{i+1}} z(s_i,s_{i+1}) \beta_{i+1}^{(z)}(s_{i+1}) \\
\beta_i^{(h)}(s_i)=&\sum_{s_{i+1}} z(s_i,s_{i+1}) \ \cdot \nonumber\\
&\big(\beta_{i+1}^{(h)}(s_{i+1}) + h(s_i,s_{i+1})
\alpha_{i+1}^{(z)}(s_{i+1})\big),
\end{align}
or equivalently
\begin{align}
\label{emp_hmm: beta_z_rec}
\beta_t^{(z)}&(i)=\sum_{j} \frac{\abHMM{i}{j}{t}}{c_{t+1}}  \beta_{t+1}^{(z)}(j) \\
\label{emp_hmm: beta_h_rec} \beta_t^{(h)}&(i)=
\sum_{j} \frac{\abHMM{i}{j}{t}}{c_{t+1}}  \ \cdot \nonumber\\
&\Big(\beta_{t+1}^{(h)}(j) + \beta_{t+1}^{(z)}(j) \log
\frac{\abHMM{i}{j}{t}}{c_{t+1}}  \Big),
\end{align}
where the normalization constants $c_t$ are computed in the
forward pass.

\subsection{HMM entropy computation using ESRFB}
\label{FB_HMM_Entropy}

If the summation of the global kernel (\ref{esrfa: u_p_plogp}) is
performed over the whole sequence
\begin{equation}
\bigoplus_{\bs s} u(\bs s) =\bigoplus_{\bs s}\ \big(\ p(\bs s | \bs o)\ , \ p(\bs s | \bs o)\log p(\bs s | \bs o) \ \big).
\end{equation}
the $z$ and $h$ parts of the sum reduce to
\begin{align}
&\uSum{z}=\sum_{\bs s} p(\bs s | \bs o) = 1,\\
&\uSum{h}=\sum_{\bs s} p(\bs s | \bs o) \log p(\bs s | \bs o)= - H(\bs S |\bs o) 
\end{align}
The $h$ part of the sum corresponds to the \textit{HMM} entropy and it can be found as a solution of the normalization problem
\begin{align}
\label{esrfa: sum_u_h}%
\uSum{h}&=\sum_{s_T}\alpha_T^{(h)}(s_T)
\end{align}
using only the forward pass, according to equations (\ref{emp_hmm:
alpha_z_init})-(\ref{emp_hmm: alpha_h_init}), (\ref{emp_hmm:
alpha_z_rec})-(\ref{emp_hmm: c_t}), as follows.

\subsubsection{Initialization}
For $j = 1,\dots,N$ set:

\begin{equation}
\label{emp_hmm: en: c_0}
c_0=\sum_{j=1}^N \pi_{j} b_{j}(o_0)\quad
\alpha_0^{(z)}(j)=\frac{\pi_{j} b_{j}(o_0)}{c_0},
\end{equation}
\begin{equation}
\alpha_0^{(h)}(j)=\frac{\pi_{j} b_{j}(o_0)}{c_0} \log
\frac{\pi_{j} b_{j}(o_0)}{c_0}.
\end{equation}

\subsubsection{Induction} For $1 \leq t \leq T$, $1 \leq j \leq N$ compute
\begin{align}
\label{emp_hmm: en: c_t}
&c_t =\sum_{j=1}^N \sum_{i=1}^N \alpha_{t-1}^{(z)}(i)  \abHMM{i}{j}{t} \\
\label{emp_hmm: en: alpha_t_z}
\alpha_t^{(z)}&(j)=\sum_{i=1}^N \frac{a_{ij} b_j(o_t)}{c_t} \cdot \alpha_{t-1}^{(z)}(i) \\
\label{emp_hmm: en: alpha_t_h}
\alpha_i^{(h)}&(j)=\sum_{i=1}^N \frac{a_{ij} b_j(o_t)}{c_t} \ \cdot \nonumber\\
&\big(\alpha_{t-1}^{(h)}(i) + \alpha_{t-1}^{(z)}(i) \log \frac{a_{ij} b_j(o_t)}{c_t} \big).
\end{align}

\subsubsection{Termination}
Terminate algorithm with summations:
\begin{align}
\label{emp_hmm: en: H_s_o}%
H(\bs S |\bs o)= - \sum_{j=1}^N\alpha_T^{(h)}(j),
\end{align}

The \emph{ESRFB} algorithm runs in $\mc O(N^2 T)$ time using $\mc
O(N)$ space as in Hernando et al.'s algorithm. Moreover, both
algorithms recursively compute the forward probability
\begin{equation}
\alpha_t^{(z)}(s_t)=p(s_t | o_{0:t}).
\end{equation}
The difference in two algorithms is in the second quantity which
is computed - in Hernando et al.'s algorithm it is the
intermediate entropy
\begin{align}
H_t(s_t)&=H(S_{0:t-1}|s_t,o_{1:t})=\nonumber \\%
&-\sum_{s_{0:t-1}}p(s_{0:t-1}|s_t,o_{1:t})%
\log p(s_{0:t-1}|s_t,o_{1:t}),
\end{align}
while in the \emph{ESRFB} it is the $h$-part of the forward
vector:
\begin{equation}
\alpha_t^{(h)}(s_t) =%
\sum_{s_{o:t-1}} p(s_{0:t} | o_{0:t}) \log p(s_{0:t} | o_{0:t})
\end{equation}
The relation between the quantities
\begin{equation}
\label{emp_hmm: en: alpha_Ht}%
\alpha_t^{(h)}= \alpha_t^{(z)} H_t(s_t) + \alpha_t^{(z)} \log
\alpha_t^{(z)}
\end{equation}
can easily be derived by use of the elementary probability
transformations.

Furthermore, from \emph{HMM} joint probability factorization
(\ref{HMM p expanded}) we can derive the Markov properties
\begin{align}
&p(o_t | s_t, s_{t-1}, o_{0:t-1})=p(o_t | s_t)\\
&p(s_t | s_{t-1}, o_{0:t-1})=p(s_t | s_{t-1}),
\end{align}
which imply the following equalities:
\begin{multline}
\label{emp_hmm: en: z}
\frac{a_{s_{t-1}s_t} b_{s_t}(o_t)}{c_t} =
\frac{p(s_t|s_{t-1}) p(o_t|s_t)}{p(o_t | o_{0:t-1})}=\\
\frac{p(o_t, s_t | s_{t-1}, o_{0:t-1})}{p(o_{t}| o_{0:t-1})} =
\frac{p_{t-1|t}(s_{t-1}|s_t) \cdot
\alpha_t^{(z)}(s_t)}{\alpha_{t-1}^{(z)}(s_{t-1})}.
\end{multline}
where $\hat p_{t-1|t}(s_{t-1}|s_t) = p(s_{t-1}|s_t, o_{0:t})$ as
defined in the Hernando et al.'s algorithm. Then, the recursive
equations for $H_t(s_t)$ derived by Hernando et al. can also be
obtained from the \emph{ESRFB} algorithm by substituting
(\ref{emp_hmm: en: z}) and (\ref{emp_hmm: en: alpha_Ht}) in
recursive equations for $\alpha_t^{(h)}$ in \emph{ESRFB}
algorithms, which give us the close relation between two
algorithms.

\subsection{HMM subsequence constrained entropy computation using ESRFB}
If the summation of the global kernel is performed over a subsequence $s_{-l:r}$
\begin{equation}
\bigoplus_{s_{-l:r}} u(\bs s) =\bigoplus_{s_{-l:r}}\ \big(\ p(\bs s | \bs o)\ , \ p(\bs s | \bs o)\log p(\bs s | \bs o) \ \big).
\end{equation}
the $z$ and $h$ parts of the sum are
\begin{align}
\Big(\bigoplus_{s_{-l:r}} u(\bs s) \Big)^{(z)} &=p(s_{l:r}),\\ 
\Big(\bigoplus_{s_{-l:r}} u(\bs s) \Big)^{(h)}
&=- H(S_{l:r}, s_{l:r} |\bs o). 
\end{align}
The $h$ part of the sum corresponds to the \textit{HMM} subsequence constrained entropy and it can be found as a solution of the marginalization problem
\begin{equation}
v_{l:r}(s_{l:r})=
\alpha_l(s_l) \otimes \bigotimes_{i=l+1}^r u_i(s_{i-1}, s_i)
 \otimes \beta_r(s_r).
\end{equation}
The $z$ and $h$ parts can be found using the definition for the
entropy semiring operations:
\begin{equation}
\label{emp_hmm: suben: v_lr_z}
\Big(\bigoplus_{s_{- l:r}} u(\bs s)\Big)^{(z)}=
\alpha_l^{(z)}(s_l) \beta_r^{(z)}(s_r)
\prod_{i=l+1}^r z_i(s_{i-1}, s_i),
\end{equation}

\begin{multline}
\label{emp_hmm: suben: v_lr_h}
\Big(\bigoplus_{s_{- l:r}} u(\bs
s)\Big)^{(h)}=
\prod_{i=l+1}^r z_i(s_{i-1}, s_i) \cdot \\
\big( \alpha_l^{(z)}(s_l) \beta_r^{(h)}(s_r) +
\alpha_l^{(h)}(s_l) \beta_r^{(z)}(s_r) + \\
\alpha_l^{(z)}(s_l) \beta_r^{(z)}(s_r) \sum_{j=l+1}^r h_j(s_{j-1}, s_j)\big)
\end{multline}
To compute the $h$-part of the marginal, we need $l$-th forward
and $r$-th backward vectors. The $l$-th forward vector can be
computed by \emph{ESR} forward algorithm using recursive equations
(\ref{emp_hmm: en: c_0})-(\ref{emp_hmm: en: alpha_t_h}). However,
the recursive steps (\ref{emp_hmm: en: c_t})-(\ref{emp_hmm: en:
alpha_t_h}) for the normalization constants $c_t$ and the $z$ part
of forward vectors should be performed for all $t$, because the
normalization constants $c_t, r< t \leq T$ should be available in
the backward pass. Once the normalization constants are computed,
the backward pass can be performed according to the equations
(\ref{emp_hmm: beta_z_init}), (\ref{emp_hmm: beta_z_rec})-(\ref{emp_hmm: beta_h_rec}), and,
after that, we can compute the subsequence constrained entropy
using the equalities (\ref{emp_hmm: suben: v_lr_z})-(\ref{emp_hmm:
suben: v_lr_h}) and (\ref{hmm_en: H_S_lr_final}). The algorithm
follows.

\subsubsection{Forward initialization}
For $j = 1,\dots,N$ set:
\begin{equation}
c_0=\sum_{j=1}^N \pi_{j} b_{j}(o_0)\quad
\alpha_0^{(z)}(j)=\frac{\pi_{j} b_{j}(o_0)}{c_0}, 
\end{equation}
\begin{equation}
\alpha_0^{(h)}(j)=\frac{\pi_{j} b_{j}(o_0)}{c_0} \log
\frac{\pi_{j} b_{j}(o_0)}{c_0}.
\end{equation}

\subsubsection{Full forward recursion} For $1 \leq t \leq l$, $1 \leq j \leq N$ compute
\begin{align}
&c_t =\sum_{j=1}^N \sum_{i=1}^N \alpha_{t-1}^{(z)}(i)  \abHMM{i}{j}{t} \\
\alpha_t^{(z)}&(j)=\sum_{i=1}^N \frac{a_{ij} b_j(o_t)}{c_t} \cdot \alpha_{t-1}^{(z)}(i) \\
\alpha_i^{(h)}&(j)=\sum_{i=1}^N \frac{a_{ij} b_j(o_t)}{c_t} \ \cdot \nonumber\\
&\big(\alpha_{t-1}^{(h)}(i) + \alpha_{t-1}^{(z)}(i) \log
\frac{a_{ij} b_j(o_t)}{c_t} \big).
\end{align}

\subsubsection{Forward $z$-part recursion} For $ l+1 \leq t \leq T$, $1 \leq j \leq N$ compute
\begin{align}
\label{emp_hmm: en: c_t}
&c_t =\sum_{j=1}^N \sum_{i=1}^N \alpha_{t-1}^{(z)}(i)  \abHMM{i}{j}{t} \\
\alpha_t^{(z)}&(j)=%
\sum_{i=1}^N \frac{a_{ij} b_j(o_t)}{c_t} \cdot
\alpha_{t-1}^{(z)}(i).
\end{align}

\subsubsection{Backward initialization}
For $j = 1,\dots,N$ set:
\begin{equation}
\beta_T^{(z)}(j)=1, \quad
\beta_T^{(h)}(j)=0.
\end{equation}

\subsubsection{Backward recursion} For $T-1 \geq t \geq r$, $1 \leq j \leq N$ compute
\begin{align}
\beta_t^{(z)}&(i)=\sum_{j} \frac{\abHMM{i}{j}{t}}{c_{t+1}}  \beta_{t+1}^{(z)}(j) \\
\beta_t^{(h)}&(i)=
\sum_{j} \frac{\abHMM{i}{j}{t}}{c_{t+1}}  \ \cdot \nonumber\\
&\Big(\beta_{t+1}^{(h)}(j) + \beta_{t+1}^{(z)}(j) \log \frac{\abHMM{i}{j}{t}}{c_{t+1}}  \Big).
\end{align}

\subsubsection{Termination} For $l\leq t \leq r$, $1\leq s_t \leq N$, compute the subsequence constrained entropy:
\begin{equation}
p(s_{l:r})=
\alpha_l^{(z)}(s_l) \beta_r^{(z)}(s_r)
\prod_{t=l+1}^r \frac{a_{s_{t-1} s_t} b_{s_t}(o_t)}{c_t},
\end{equation}

\begin{multline}
\label{emp_hmm: suben: v_lr_h}%
-H(S_{l:r}, s_{l:r} | \bs o)=
\prod_{t=l+1}^r
\frac{a_{s_{t-1} s_t} b_{s_t}(o_t)}{c_t} \cdot \\
\big( \alpha_l^{(z)}(s_l) \beta_r^{(h)}(s_r) +
\alpha_l^{(h)}(s_l) \beta_r^{(z)}(s_r) + \\
\alpha_l^{(z)}(s_l) \beta_r^{(z)}(s_r) \sum_{j=l+1}^r h_j(s_{j-1}, s_j)\big)
\end{multline}

\begin{equation}
H(S_{- l:r} | s_{l:r},  \bs o) = \frac{H(S_{- l:r} , s_{l:r}| \bs o) + \log p(s_{l:r} | \bs o )}{p(s_{l:r} | \bs o )}
\end{equation}

The time complexity of the algorithm is $\mc O(N^2 T + N^{r-l})$,
where $\mc O(N^2 T)$ is for the forward-backward recursion, and
$\mc O(N^{r-l})$ for the termination phase, which is the same time
complexity as in Mann-MacCallum's algorithm.

On the other hand, full forward recursion phase can be realized in
$\mc O(N^2 l)$ time and in fixed size memory $\mc O(N)$, since
$\alpha_{t-1}^{(z)}$, $\alpha_{t-1}^{(h)}$ and $c_{t-1}$ can be
deleted after having been used for the computation of
$\alpha_t^{(z)}$, $\alpha_t^{(h)}$ and $c_t$. Similarly, the
forward $z$-part recursion and backward pass requires $\mc O(N)$
space. Only additional space depending on the sequence length $\mc
O(T-l)$ should be available for normalization constants in the
forward $z$-part recursion phase, since they should be available
in the backward and termination phases. Finally, regarding $\mc
O(N^{r-l})$ space required for storing the results in the
termination phase, the total memory complexity is $\mc O(T-l +
N^{r-l})$, which slightly increases with $T$ then $O(N T +
N^{r-l})$, as required by Mann-MacCallum's algorithm.

\section{Conclusion}

This paper proposes a new algorithm for memory efficient
computation of the \textit{HMM} entropy and subsequence
constrained entropy when the observation sequence is given. The
algorithm is called Entropy Semiring Forward-backward
(\emph{ESRFB}) since it is based on forward-backward recursion
over the entropy semiring in the same manner as in our previous
paper \cite{Ilic_et_al_11}.

\emph{ESRFB} has the same time complexity as a previously
developed algorithm for subsequence constrained \emph{HMM} entropy
computation developed by Mann and MacCallum
\cite{Mann_McCallum_07}, but with lower memory requirements. It is
also applicable to state sequence entropy computation running with
the same time and memory complexity as the recursive algorithm
proposed by Hernando et al. \cite{Hernando_et_al_05}. In addition,
we have shown how the recursive equations in Hernando et al.'s
algorithm can be derived from the \emph{ESRFB} recursive
equations.

\bibliographystyle{plain}

\end{document}